\documentclass[aps,prb,twocolumn,showpacs,floatfix]{revtex4}

\usepackage{graphics}

\begin{document}

\widetext

\title{ Remarkable site selectivity properties of the ARPES matrix
  element in Bi$_2$Sr$_2$CaCu$_2$O$_8$}

\author{S. Sahrakorpi$^{1}$, M. Lindroos$^{1,2}$ and A. Bansil$^1$}
\affiliation{$^1$Physics Department, Northeastern University, Boston,
  Massachusetts 02115 \\
  $^2$Institute of Physics, Tampere University of Technology, 
  P.O. Box 692, 33101 Tampere, Finland}

\date{\today}

\begin{abstract}
  
  We show that the ARPES spectra for emission from the bonding as well
  as the antibonding Fermi surface sheet in Bi$_2$Sr$_2$CaCu$_2$O$_8$
  (Bi2212) possess remarkable site selectivity properties in that the
  emission for photon energies less than 25 eV is dominated by $p
  \rightarrow d$ excitations from just the O-sites in the CuO$_2$
  planes. There is little contribution from Cu electrons to the ARPES
  intensity, even though the initial states at the Fermi energy
  contain an admixture of Cu-$d$ and O-$p$ electrons.  We analyze the
  origin of this effect by considering the nature of the associated
  dipole matrix element in detail and find that various possible
  transition channels (other than $p \rightarrow d$ on O-sites) are
  effectively blocked by either the fact that the related radial cross
  section is small and/or a lack of available final states. Our
  prediction that ARPES can preferentially sample Cu or O states by
  tuning the photon energy suggests novel possibilities for exploiting
  energy dependent ARPES spectra for probing initial state characters
  in the cuprates.
 
\end{abstract}


\maketitle 

\section{Introduction}

The photointensity observed in an angle-resolved photoemission (ARPES)
experiment is the result of modulation of the spectral density of the
initial state via the ARPES matrix element. This modulation can be
quite substantial in the cuprates and depends in general strongly on
${\mathbf k}_\parallel$, polarization and energy of the incident
photons as well as the energy and character of the initial
state.\cite{bansil99,lindroos02,bansil02,asensio03,chuang03,%
  feng02,kordyuk02,damascelli02,campuzano02} Chuang et
al.\cite{chuang03}, for example, have recently exploited the large
theoretically predicted differences in the ARPES cross sections of
bonding and antibonding pieces of the Fermi surface (FS) by tuning the
photon energy to adduce the doping dependence of the bilayer splitting
in Bi$_2$Sr$_2$CaCu$_2$O$_8$ (Bi2212).  Asensio et al.\cite{asensio03}
find that apparently different looking Fermi surface maps obtained in
Bi2212 over wide area in the $({\mathbf k}_x,{\mathbf k}_y)$ plane at
different photon energies largely reflect the effect of the ARPES
matrix element. In other applications, such as a recent
proposal\cite{vekhter02} to determine the spectral function of the
boson mediating the Cooper pairs in the cuprates, it will be necessary
to develop strategies for minimizing the effect of the ARPES matrix
element in order to obtain robust physical results through ARPES
spectra. It is clear that a good understanding of the nature of the
ARPES matrix element is of considerable importance in continued
development of ARPES as a probe of electronic structure of complex
materials.
 
With this motivation, we examine in this article the nature of
emissions from the bonding as well as the antibonding parts of the FS
of Bi2212.  In considering contributions to photointensity arising
from different angular momentum channels and various atomic sites in
the unit cell, we find that the ARPES matrix element possesses
remarkable site selectivity properties in that the ARPES intensity
throughout the $({\mathbf k}_x,{\mathbf k}_y)$ plane in the 5-25 eV
energy range is dominated by only the O $p \rightarrow d$ transitions.
The contribution from Cu sites is quite small, even though the
electronic states at the Fermi energy in Bi2212 contain substantial Cu
character.  In order to ascertain the origin of this effect, the
dipole transition matrix element between the relevant initial and
final states is analyzed in detail and the factors responsible for
this behavior are identified.\footnote{The site selectivity effect
  discussed in this article was noted in Ref.~\onlinecite{lindroos02}.
  However, Ref.~\onlinecite{lindroos02} considered only a single
  ${\mathbf k}_\parallel$-point, namely the $M (\pi,0)$ symmetry point
  and did not attempt to elucidate the origin of this effect in terms
  of radial cross sections, etc. as done in this study.}  Our
prediction that ARPES can preferentially sample Cu or O states
suggests novel possibilities for exploiting energy dependent ARPES
spectra to gain insight into character of initial states in complex
materials.

The paper is ordered as follows. In Section II we briefly discuss the
underlying formalism.  Section III presents and analyzes the nature of
the ARPES spectra. In Subsection IIIA, a typical ARPES spectrum for
emission from the Fermi energy in Bi2212 is considered to orient the
reader.  Subsection IIIB continues with a discussion of site-resolved
photointensities over the 5-25 eV photon energy range for ${\mathbf
  k}_\parallel$ values covering the entire $({\mathbf k}_x,{\mathbf
  k}_y)$ plane. The origin of the site-selectivity effect is then
delineated in Subsection IIIC.  Finally, Section IV makes some
concluding remarks, including possible application of site-selectivity
property in gaining insight into interesting issues such as the
Zhang-Rice mechanism\cite{zhang88,harada02} in the cuprates and the
doping dependence of the Hubbard $U$ parameter via energy dependent
ARPES measurements\cite{kusko02}.
 
\section{Overview of Formalism}

The methodology used in this study has been discussed previously in
Ref.~\onlinecite{lindroos02}, to which we refer the reader for
details; See also Refs. \onlinecite{gofron94,lindroos95,bansil95,%
  lindroos96,bansil99,sahrakorpi01,sahrakorpi02}.  Some comment in
this connection is nevertheless necessary in order to meaningfully
describe the new results presented here. In particular,
Ref.~\onlinecite{lindroos02} shows that substantial insight into the
nature of the ARPES photointensity\footnote{This bulk momentum matrix
  element of course does not take into account effects of surface
  termination, the finite lifetimes of the initial and final states,
  etc.}  resulting from excitation between specific bulk initial and
final states in the solid, $\tilde{\psi}_i$ and $\tilde{\psi}_f$, can
be obtained in terms of the behavior of the corresponding momentum
matrix element, \mbox{$< {\tilde{\psi}_f}| \> {\bf p} \>
  |{\tilde{\psi}_i}>$}.  We then expand the $\tilde{\psi}$ within the
KKR band structure scheme\cite{mijnarends90,bansil99_2} as
\begin{equation}
  \tilde{\psi}({\mathbf r}) =  
  \sum_{L,\beta} \> i^l C_L^\beta R_l^\beta (r) Y_L(\Omega)
  \label{y8} \>,
\end{equation}
where $L \equiv (l,m)$ is a composite angular momentum index,
$Y_L(\Omega)$ are real spherical harmonics and $\beta$ denotes
different basis sites. $C_L^\beta$ are expansion coefficients and
$R_l^\beta (r)$ is the radial part of the Bloch wave function on site
$\beta$. The use of form~(\ref{y8}) for the initial and final states
yields\cite{chen76,sahrakorpi01,lindroos02,sahrakorpi02}
\begin{equation}
  <\tilde{\psi}_f|{\bf p}|\tilde{\psi}_i> = 
  \sum_{\alpha ,\beta} \sum_{L,L'} 
  {\mathbf {\hat{e}}}_\alpha i^{l-l'-1} {C_{L'}^\beta}^* C_L^\beta  
  B_{l,l'}^\beta {\cal G}_{L,L'}^\alpha
  \label{y9} \>.
\end{equation}
Here ${\mathbf {\hat{e}}}_\alpha$ denotes a unit vector along
$\alpha$-direction and primed indices refer to the final state.
${\cal G}_{L,L'}^\alpha$ are the standard Gaunt coefficients.  The
detailed expression for $B_{l,l'}^\beta$ is given in Ref.
\onlinecite{lindroos02}. For present purposes, the important point to
recognize is that
\begin{equation}
  B_{l,l'}^{\beta} \propto \int_0^{r_{MT}^\beta}
  ~{R_{l'}^{\beta}(r)}^* ~r~ R_l^{\beta}(r) ~r^2~ dr
  \label{y10} \>,
\end{equation}
where the integral extends to the muffin-tin radius $r_{MT}^\beta$ of
the atom $\beta$ in the basis.  It is useful to decompose the momentum
matrix element into contributions from various angular momentum
channels and sites in the unit cell as follows:
\begin{eqnarray}
  M_\alpha \> \equiv \> 
  < \> \tilde{\psi}_f \> | \> p_\alpha \> | \> \tilde{\psi}_i \> >
  & = & \> \sum_{\gamma} \> M^\gamma_\alpha  
  \label{y12} \\
  & = & \> \sum_{\gamma} \> \sum_{L,L'} M^{\gamma}_{\alpha, \> L,L'}  
  \label{y13} \>,
\end{eqnarray}
with 
\begin{equation}
  M^{\gamma}_{\alpha, \> L,L'} \> = \> \sum_{\delta} \> 
  i^{l-l'-1} \> ({C_{L'}^{\gamma,\delta}})^* \> 
  C_L^{\gamma,\delta} \> B_{l,l'}^{\gamma,\delta} \> {\cal G}_{L,L'}^\alpha 
  \label{y15} \>.
\end{equation}
Aside from the obvious notation in Eqs.~\ref{y12}-\ref{y15}, indices
$\gamma$ and $\delta$ are together meant to encompass the summation of
Eq.~\ref{y9} over all basis sites $\beta$ in the unit cell.  In the
present case of tetragonal Bi2212, $\gamma$ takes on eight distinct
values which include the Ca atom and seven pairs of symmetrically
located atoms (Bi, O$_{Bi}$, Sr, O$_{Sr}$, Cu, O$_{Cu,x}$,
O$_{Cu,y}$).  $\delta$ in Eq.~\ref{y15}, on the other hand, takes only
two values, which account for the two atoms placed symmetrically with
respect to the Ca layer in the unit cell.  In this sense, we may think
of $\gamma$ as a ``site" index and $\delta$ as a "pairing" index.  In
this work we study particularly the matrix element
$M^{\gamma}_{\alpha, \> L,L'}$, which includes the contribution of
pairs of sites $\delta$ related by mirror symmetry. We will see that
its dependence on different sites and excitation channels $L
\rightarrow L'$ is important for delineating the nature of
photointensity in Bi2212.

Concerning relevant computational details, the crystal potential used
is the same as that employed in our previous studies of Bi2212 on a
body centered tetragonal lattice and involves 30 atoms per unit cell;
See, e.g., Refs. \onlinecite{bansil99} and~\onlinecite{lindroos02}.
For definiteness, incident light is assumed polarized along the
[110]-direction. The damping of final states is included via an
imaginary part of the final state self-energy, $\Sigma_f'' = 2$ eV.
In order to understand the ${\mathbf k}_\parallel$ and energy
dependencies of the dipole matrix element, calculations were carried
out over the photon energy range of 5-25 eV for each of the 20
${\mathbf k}_\parallel$-points considered in the $({\mathbf
  k}_x,{\mathbf k}_y)$-plane.  This set of ${\mathbf
  k}_\parallel$-points covers both the bonding and the antibonding
parts of the bilayer split Fermi surface of Bi2212.

\section{Results}

\subsection{ARPES spectrum of Bi2212}

\begin{figure}
  \begin{center}
    \resizebox{6.5cm}{!}{\includegraphics{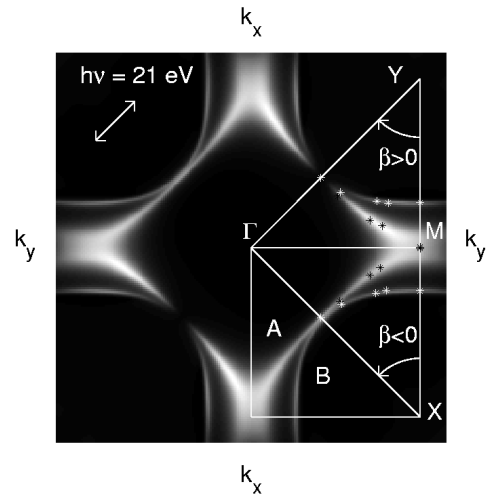}}
  \end{center}             
  \caption{
    Grey scale plot (white denotes high) of theoretical ARPES
    intensity for emission from the Fermi energy at $h\nu=21$ eV for
    tetragonal Bi2212. Light is assumed to be polarized along $[110]$
    (white arrow).  White and black stars give the ${\mathbf
      k}_\parallel$-points (corresponding to different angles $\beta$)
    where the energy dependence of the dipole matrix element for the
    bonding (B) and antibonding (A) Fermi surface sheets shown in
    Figs.~\ref{fig2} and~\ref{fig3} has been computed.}
\label{fig1}               
\end{figure}
For orientation, Fig.~\ref{fig1} shows a typical computed one-step
ARPES spectrum for emission from the $E_F$ at 21 eV.  The Umklapp and
shadow features\cite{bansil02,asensio03} are not included in this
simulation for simplicity.  The imprints of the bilayer-split bonding
(B) and antibonding (A) parts of the FS are seen clearly and are in
remarkable accord with recent ARPES
experiments\cite{bogdanov01,kordyuk02,bansil02,asensio03,chuang03}.
The bonding band gives rise to the $X \> (Y)$ centered hole sheet B.
The antibonding band is responsible for the electron-like
$\Gamma$-centered sheet A with a very narrow neck around the
$M$-point\cite{lindroos02}.  The associated intensity appears somewhat
diffuse due in part to the presence of a van Hove singularity in the
electronic spectrum close to $M$. Note that the polarization vector of
the incident light (white arrow) breaks the symmetry between the
irreducible regions $\Gamma-Y-M$ and $\Gamma-M-X$ of the Brillouin
zone. For this reason, we have characterized the ${\mathbf k}$-points
on the A and B FS's in these two regions by the indicated angle
$\beta$, where $\beta>0$ refers to the upper half and $\beta<0$ to the
lower half on the right hand side of the figure. The intensity from
either the A or the B sheet is zero along the nodal line $\Gamma-Y$
$(\beta=45^\circ)$, but this is not the case for the symmetrically
placed nodal line $\Gamma-X$ for $\beta=-45^\circ$. This is due to the
fact that even though the momentum matrix elements $M_x$ and $M_y$ are
non-zero in general they add (for the present polarization)
destructively along $\Gamma-Y$ but constructively along the $\Gamma-X$
line.

\begin{figure}
  \begin{center}
    \resizebox{8.5cm}{!}{\includegraphics{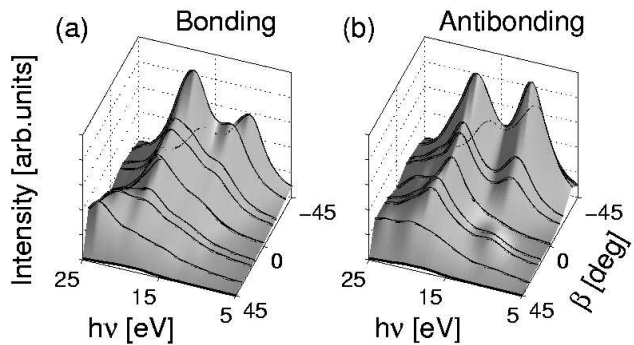}}
  \end{center}             
  \caption{
    Intensity $\tilde{I}$ (Eq.~\ref{tildeI}) for ${\mathbf
      k}_\parallel$-points (given by the angle $\beta$ of
    Fig.~\ref{fig1}) lying along the bonding and antibonding Fermi
    surface sheets in tetragonal Bi2212 over the photon energy range
    of $h\nu=5-25$ eV.  Intensities have been smoothed to reflect
    final state damping effect.  Results in (a) and (b) are not
    normalized in relation to each other.}
  \label{fig2}               
\end{figure}

As already noted, insight into the ARPES intensity can be obtained by
considering the momentum matrix element \mbox{$< {\tilde{\psi}_f}| \>
  {\bf p} \> |{\tilde{\psi}_i}>$} which connects the relevant initial
and final states and the associated intensity
\begin{equation}
  \tilde{I} \> = \> { | \mathbf{A} \cdot < {\tilde{\psi}_f}|
    \mathbf{p}|{\tilde{\psi}_i}> | }^2 
  \label{tildeI} \>.
\end{equation}
Fig.~\ref{fig2} shows intensity computed by using Eq.~\ref{tildeI} for
a series of ${\mathbf k}_\parallel$-points lying along the bonding and
antibonding Fermi surface sheets characterized by the angle $\beta$ of
Fig.~\ref{fig1}. It is clear that the dipole matrix element varies
strongly with photon energy and ${\mathbf k}_\parallel$.  The main
features in the intensity from both the bonding and antibonding bands
are the two peaks at $h\nu$ around 11 eV and 18 eV.

\subsection{Contributions to photointensity from different sites}

\begin{figure}
  \begin{center}
    \resizebox{8.5cm}{!}{\includegraphics{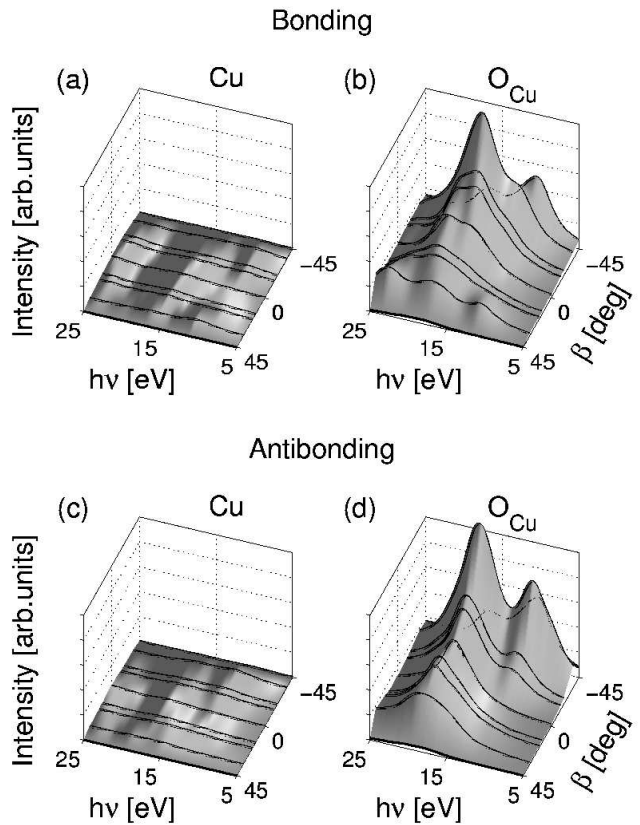}}
  \end{center}             
  \caption{
    Contributions to the photointensities of Fig.~\ref{fig2} from Cu
    and O sites in the CuO$_2$ planes in the crystal lattice. (a) and
    (c) give contribution from Cu sites for transitions from the
    bonding and the antibonding initial states respectively. (b) and
    (d) similarly refer to the O-sites. See caption to Fig.~\ref{fig2}
    for other details. }
  \label{fig3}               
\end{figure}  
We next consider the extent to which Cu and O related initial state
electrons are excited in the photoemission process. Fig.~\ref{fig3}
gives site ($\gamma$) resolved contributions to the photointensities
obtained by taking into account only the terms related to Cu or
O$_{Cu}$-sites of the CuO$_2$ planes, i.e.  for $\gamma\equiv$ Cu or
$\gamma\equiv$ O$_{Cu,x}$+O$_{Cu,y}$ in Eq.~\ref{y12}, where the
corresponding contributions from the two CuO$_2$ planes are added
together.  Fig.~\ref{fig3}(a) shows the Cu contribution from the
bonding initial state as a function of ${\mathbf k}_\parallel$ and
photon energy. The intensity is seen to be practically zero indicating
that, for emission emerging from any part of the bonding FS, Cu sites
essentially do not contribute to the photointensity.  In sharp
contrast, contribution from O$_{Cu}$-sites in Fig.~\ref{fig3}(b) is
substantial. In fact, the remaining sites in the crystal lattice (not
shown) are found to give little contribution, which explains why the
intensity pattern from O$_{Cu}$-sites in Fig.~\ref{fig3}(b) is quite
similar to the corresponding total intensity in Fig.~\ref{fig2}(a).
Much of the preceding commentary concerning the bonding FS is also
seen to be valid for the antibonding FS by looking at
Figs.~\ref{fig3}(c),~\ref{fig3}(d) and~\ref{fig2}(b) and need not be
repeated.  The results of Fig.~\ref{fig3} thus substantially extend
the observations made in Ref.~\onlinecite{lindroos02} at a single
${\mathbf k}_\parallel$-value (i.e.  $M(\pi,0)$ point) to show that
the ARPES matrix element is dominated by contribution from
O$_{Cu}$-sites in the CuO$_2$ planes over the {\em entire FS} at
photon energies below 25 eV.  We have carried out limited computations
at higher energies and find that around $h\nu = 40$ eV both Cu and
O$_{Cu}$ contribute roughly equally to the photointensity.

\subsection{Why copper electrons do not contribute?}

Eqs.~\ref{y9} and~\ref{y15} make it clear that the value of the
momentum matrix element depends on the magnitudes and phases of
several component quantities, namely, the expansion coefficients
$C_L^{\gamma,\delta}$ of the initial and final states, the Gaunt
coefficients ${\cal G}_{L,L'}^\alpha$, and the radial integral terms
$B_{l,l'}^{\gamma,\delta}$.  Concerning the character of the {\em
  initial states} of interest here, note that the antibonding as well
as the bonding initial states in the vicinity of the FS in Bi2212
primarily consist of electrons in the CuO$_2$ bilayers, where the
O$_{Cu,x}$, O$_{Cu,y}$ and Cu atomic sites contain approximately
80-90\% of the weight in the associated Bloch wave
function\cite{lindroos02}.  The primary Cu orbital involved is
$d_{x^2-y^2}$, while the most relevant O$_{Cu}$ orbitals are $p_x$ and
$p_y$.  The distribution of electrons between different sites and
orbitals changes as a function of ${\mathbf k}_\parallel$, but in
general the electrons at $E \approx E_F$ are strongly concentrated in
the vicinity of the CuO$_2$ planes.

\begin{figure}
  \begin{center}
    \resizebox{8.5cm}{!}{\includegraphics{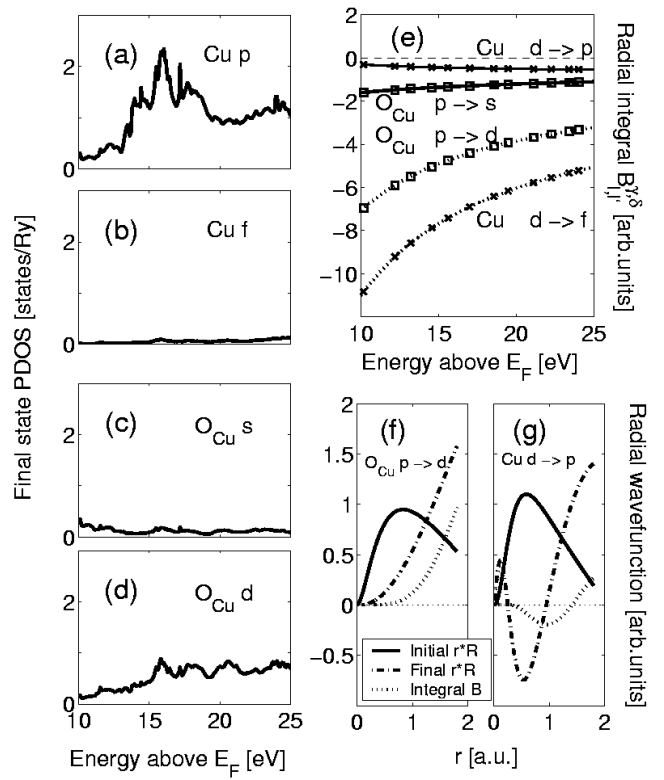}}
  \end{center}             
  \caption{
    Left-hand side: PDOS's for final states are presented as a
    function of energy above Fermi energy ($E_F$) for the most
    relevant angular momentum channels.  (a) and (b) show the PDOSs of
    Cu $p$ and $f$, and (c) and (d) show the O $s$ and $d$ final
    states.  Right-hand side: (e) gives the radial integral term
    $B_{l,l'}^{\gamma,\delta}$ of Eq.~\ref{y15} for indicated $l
    \rightarrow l'$ transitions from the bonding initial state at
    $\beta=0^\circ$ (near $M$-point) as a function of final state
    energy.  The data with crosses (squares) is for Cu (O$_{Cu}$)
    sites of the CuO$_2$ planes. In (f) and (g), typical initial
    (solid) and final (dash-dotted) state radial wave functions
    $R_{l'}^{\beta}(r)$ and $R_l^{\beta}(r)$, respectively, are shown
    as a function of radius $r$ for O$_{Cu}$ $p \rightarrow d$ and Cu
    $d \rightarrow p$ transitions, respectively.  The dotted line
    shows the behavior of the radial overlap integral
    (Eq.~\ref{y10}).}
  \label{fig4} 
\end{figure}      
Since the initial states of interest are mainly Cu $d$ and O$_{Cu}$
$p$ type, from various possible $L\rightarrow L'$ transition channels
in Eq.~\ref{y15}, the dipole selection rule $\Delta l = \pm 1$ limits
the $L'$ for the {\em final states} to the Cu $p$ and $f$ states and
O$_{Cu}$ $s$ and $d$ states, respectively.  A handle on the
availability of appropriate final states (i.e. a substantial final
state coefficient $C_{L'}^{\gamma,\delta}$) is provided by the partial
densities of states (PDOS's) of the final states resolved into
different $l'$-channels and sites, shown in panels (a)-(d) of
Fig.~\ref{fig4}. In the case of Cu, Figs.~\ref{fig4}(a)
and~\ref{fig4}(b) show that PDOS is substantial only in the $p$
channel, the value in the $f$ channel being quite small, so that
significant contribution to the matrix element is to be expected only
from the $d \rightarrow p$ transitions.  Similarly,
Figs.~\ref{fig4}(c) and~\ref{fig4}(d) show that for the
O$_{Cu}$-related final states, the $d$ PDOS is substantial so that the
$p\rightarrow d$ transitions from O$_{Cu}$-sites will dominate.  In
short, the lack of Cu $f$-type and O$_{Cu}$ $s$-type final states in
effect ``blocks'' the Cu $d \rightarrow f$ and O$_{Cu}$ $p \rightarrow
s$ channels in the allowed dipole transitions such that only Cu $d
\rightarrow p$ and O$_{Cu}$ $p \rightarrow d$ excitations remain to be
of interest in further analysis.

We next consider the term \mbox{$B_{l,l'}^{\gamma,\delta}$} of
Eq.~\ref{y15} which is defined by the radial integral of
Eq.~\ref{y10}. This term may be thought of as the radial cross section
associated with site $(\gamma,\delta)$ for the transition from $l$ to
$l'$ orbital. In Fig.~\ref{fig4}(e) these radial integrals are
presented for the relevant Cu and O$_{Cu}$ related transitions.  All
curves are seen to be smooth, slowly varying functions of energy,
although absolute values vary greatly between different transitions.
The Cu $d \rightarrow f$ and O$_{Cu}$ $p \rightarrow d$ excitation
channels are the strongest, whereas the Cu $d \rightarrow p$ and
O$_{Cu}$ $p \rightarrow s$ integrals are smaller and will attenuate
the related excitations.  In view of the discussion of the preceding
paragraph, weakness of the Cu $d \rightarrow p$ radial cross section
is particularly important, since the Cu $d \rightarrow p$ channel is
the only Cu transition channel that could have contributed
substantially to the photointensity. It follows then that the O$_{Cu}$
$p \rightarrow d$ is the main transition channel in the formation of
the momentum matrix element for emission from the FS in Bi2212 at low
photon energies.\footnote{P. Marksteiner et al., Phys. Rev. B {\bf
    38}, 5098 (1988), considered angle-integrated photointensities and
  found that the spectra are weighted in favor of Cu excitations at
  higher photon energies.}

Insight into the relative magnitudes of the O$_{Cu}$ \mbox{$p
  \rightarrow d$} and Cu \mbox{$d \rightarrow p$} related radial cross
sections can be gained by considering the behavior of initial and
final state radial wave functions as shown in panels~(f) and~(g) of
Fig.~\ref{fig4}. In the case of O$_{Cu}$ $p \rightarrow d$ transition,
initial and final states in Fig.~\ref{fig4}(f) are both nodeless and
thus the integral $B_{l,l'}^{\gamma,\delta}$ does not change sign.  By
contrast, the wave function for the Cu $p$ final state in
Fig.~\ref{fig4}(g) (dot-dashed) possesses two nodes so that when it is
combined with the nodeless Cu $d$ initial state radial wave function
the integral oscillates in sign yielding a small value of the radial
cross section for this transition.  This mechanism where the nodes in
the final state wave function play an important role is reminiscent of
the arguments made in connection with the origin of the Cooper
minima\cite{cooper62,yeh85,molodtsov00} observed in photoionization
cross sections of atomic systems at higher photon
energies.\footnote{According to J. J. Yeh and I. Lindau,
  Ref.~\onlinecite{yeh85}, the atomic photoionization cross section is
  approximately one third smaller for the Cu $3d$ than for the O $2p$
  atomic levels at photon energies around 30 eV.}  Putting it all
together, the discussion of this section shows that even though the
initial states at the FS in Bi2212 possess considerable Cu $d$
character, these states contribute little to photointensity for $h\nu
\leq 25$ eV due to a combination of two effects: The $d$ to $f$
channel is suppressed due to the lack of available $f$-type final
states, while the $d$ to $p$ channel possesses a weak radial cross
section.\footnote{The destructive interference discussed in
  Ref.~\onlinecite{lindroos02} between Cu terms of different
  CuO$_2$-planes decreases the Cu-related intensity further. This
  destructive interference seems to be more prevalent for the Cu terms
  than for the O$_{Cu}$-terms.}

\section{Summary and Conclusions}

We consider the nature of ARPES spectra from Bi2212 for emission from
the Fermi energy. Over the 5-25 eV photon energy range, the spectra
are shown to display remarkable spatial selectivity properties in the
sense that emissions are dominated by excitation from just the O-sites
in the CuO$_2$ planes, even though the initial state wave functions
involved possess substantial Cu character. This selectivity applies to
the bonding as well as the antibonding FS sheet and holds throughout
the $({\mathbf k}_x,{\mathbf k}_y)$ plane.  Insight into the origin of
this effect is obtained by considering the dipole matrix element for
transitions between the relevant initial and final states and by
decomposing this matrix element into contributions arising from
individual sites in the unit cell in various angular momentum
channels. Our analysis reveals that, of the two possible channels for
exciting Cu $d$-electrons ($d \rightarrow f$ and $d \rightarrow p$),
the $d \rightarrow f$ channel is suppressed by lack of available
$f$-type final states, while the $d \rightarrow p$ channel is
effectively blocked due to a small radial cross section (i.e. the term
$B_{l,l'}^{\gamma,\delta}$ in Eq.~\ref{y15}). Similarly, of the two
possible channels for O $p$-electrons ($p \rightarrow s$ and $p
\rightarrow d$), the $p \rightarrow s$ channel is suppressed since few
$s$-type final states are available, leaving significant contribution
to the spectrum from only the O $p \rightarrow d$ transitions.  It
turns out that the presence of nodes in the final state wave function
tends to reduce the radial cross section, which is reminiscent of
Cooper minimum type effect in photoionization cross section of atomic
systems. Limited computations at higher photon energies show that the
contribution from Cu sites increases at higher energies, and that by
around 40 eV, the Cu and O states are excited roughly equally in the
spectrum.

Our prediction that ARPES can preferentially sample Cu or O states by
tuning the photon energy suggests novel possibilities for exploiting
energy dependent ARPES spectra for probing initial state characters.
An exciting example in the cuprates would be to test the Zhang-Rice
mechanism\cite{zhang88,harada02} and potentially to deduce
experimentally the value of the Hubbard $U$ parameter as a function of
doping.  Zhang and Rice noted that large $U$ greatly restricts double
occupancy of Cu orbitals, so that the first holes doped at near half
filling -- equivalently, the states at the top (high energy part) of
the lower Hubbard band -- should have strong O character.  Thus, even
though ARPES cannot see the upper Hubbard band (and hence cannot
measure $U$ directly), it should be possible to adduce $U$ by
measuring the relative Cu/O character of states along the lower
Hubbard band, and comparing this to predictions of appropriate model
computations.  A direct determination of the doping dependence of $U$
would be an important confirmation of recent results and
predictions\cite{kusko02}.

\begin{acknowledgments}
  
  It is a pleasure to acknowledge important discussions with Bob
  Markiewicz. This work is supported by the US Department of Energy
  contract DE-AC03-76SF00098, and benefited from the allocation of
  supercomputer time at NERSC, Northeastern University's Advanced
  Scientific Computation Center (ASCC) and the Institute of Advanced
  Computing (IAC), Tampere.  One of us (S.S.) acknowledges Suomen
  Akatemia and Vilho, Yrj\"o ja Kalle V\"ais\"al\"an Rahasto for
  financial support.

\end{acknowledgments}

\end{document}